%% file: main.tex
\documentclass[10pt,twocolumn,letterpaper]{article}

\usepackage{cvpr}              %

\usepackage{graphicx}
\usepackage{amsmath}
\usepackage{amssymb}
\usepackage{booktabs}
\usepackage{multirow}
\usepackage{multicol}
\usepackage[accsupp]{axessibility}

\usepackage[pagebackref,breaklinks,colorlinks]{hyperref}

\usepackage[capitalize]{cleveref}
\crefname{section}{Sec.}{Secs.}
\Crefname{section}{Section}{Sections}
\Crefname{table}{Table}{Tables}
\crefname{table}{Tab.}{Tabs.}

\def\cvprPaperID{53} %
\def\confName{CVPR}
\def\confYear{2022}

\newcommand{\redbf}[1]{{\textbf{\color{red}{#1}}}} %
\newcommand{\blueud}[1]{{\underline{\color{blue}{#1}}}} %

\begin{document}

\title{\vspace{-2cm} VFHQ: A High-Quality Dataset and Benchmark\\ for Video Face Super-Resolution}

\author{
	Liangbin Xie\thanks{Liangbin Xie is an intern in ARC Lab, Tencent PCG.} ~$^{1,2,3}$  \hspace{9pt}  Xintao Wang$^{3}$ \hspace{9pt} Honglun Zhang$^{3}$ \hspace{9pt}  Chao Dong\thanks{Corresponding author.}~$^{1}$ \hspace{9pt} Ying Shan$^{3}$ \\
	\small{$^{1}$Shenzhen Key Lab of Computer Vision and Pattern Recognition,} \\ \small{Shenzhen Institute of Advanced Technology, Chinese Academy of Sciences} \\
	\small{$^{2}$University of Chinese Academy of Sciences} 
	\small{$^{3}$ARC Lab, Tencent PCG} \\
	{\tt\small \{lb.xie, chao.dong\}@siat.ac.cn  \{xintaowang, honlanzhang, yingsshan\}@tencent.com}\\
}

\maketitle

\begin{abstract}
   Most of the existing video face super-resolution (VFSR) methods are trained and evaluated on VoxCeleb1, which is designed specifically for speaker identification and the frames in this dataset are of low quality. 
   As a consequence, the VFSR models trained on this dataset can not output visual-pleasing results.
   In this paper, we develop an automatic and scalable pipeline to collect a high-quality video face dataset (VFHQ), which contains over $16,000$ high-fidelity clips of diverse interview scenarios. 
   To verify the necessity of VFHQ, we further conduct experiments and demonstrate that VFSR models trained on our VFHQ dataset can generate results with sharper edges and finer textures than those trained on VoxCeleb1. 
   In addition, we show that the temporal information plays a pivotal role in eliminating video consistency issues as well as further improving visual performance.
   Based on VFHQ, by analyzing the benchmarking study of several state-of-the-art algorithms under bicubic and blind settings.
\end{abstract}

\input{sections/introduction.tex}
\input{sections/related_works.tex}

\input{sections/dataset_description.tex}

\input{sections/dataset_collection.tex}

\input{sections/necessity.tex}

\input{sections/benchmark.tex}

\section{Conclusion}
Compared against high-quality face image datasets, the poor quality of training and testing video face datasets has restricted the development of multi-frame face SR research. To fill the gap between the image face dataset and video face dataset, we propose an automatic and scalable pipeline to collect high-quality face clips from web videos, and construct a Video Face dataset with High Quality (VFHQ). Based on VFHQ, we further reveal its importance for multi-frame face SR by exploring the necessity of VFHQ compared to VoxCeleb1 and FFHQ. In addition, we conduct benchmarking studies in bicubic and blind settings. Future work includes the investigation of generative facial priors in multi-frame face SR, with the help of VFHQ. 

The proposed VFHQ may have some negative social impacts, like leaking privacy. To mitigate the influence of privacy, the selected identities are celebrities and the celebrity list comes from two public datasets~\cite{nagrani2017voxceleb,cao2018vggface2}. Users are required to read the license file provided by~\cite{cao2018vggface2} carefully before downloading the data. We sincerely hope the collected VFHQ can promote the development of face-related applications.

\noindent \textbf{Acknowledgement.} This work is partially supported by the National Natural Science Foundation of China (61906184), the Joint Lab of CAS-HK, the Shenzhen Research Program (RCJC20200714114557087), the Shanghai Committee of Science and Technology, China (Grant No. 21DZ1100100).

\newpage
{\small
\bibliographystyle{ieee_fullname}
\bibliography{egbib}
}

\end{document}

%% file: sections/introduction.tex
\section{Introduction}
\label{sec:intro}

As a special category of image super-resolution (SR)~\cite{glasner2009super,dong2014learning,lim2017enhanced}, face super-resolution (FSR) is an active research topic towards face-related applications, and has attracted increasing attention.
Face super-resolution aims at restoring high resolution (HR) face images from low-resolution (LR) observations.
Existing deep-learning-based methods mainly focus on exploiting the information of a single input image with the help of various priors, such as geometry facial priors~\cite{chen2021progressive,chen2018fsrnet,yu2018face}, reference priors~\cite{li2018learning,li2020enhanced,dogan2019exemplar,li2020blind} or generative facial priors~\cite{wang2021towards,yang2021gan}.
Thanks to the powerful capacity of convolutional neural networks (CNN) and the availability of high-quality face image datasets (\textit{e.g.}, FFHQ~\cite{karras2019style}), some recent methods~\cite{chen2021progressive,wang2021towards,yang2021gan} can restore high-quality face images with the size up to $512 \times 512$ or even $1024 \times 1024$, from distorted face inputs.

Despite the rapid development of single-frame face SR, 
a few deep-learning-based methods~\cite{jin2015robust,fang2019self,xin2020video} have tried to make progress for VFSR and their performance are even significantly inferior to the results of existing single face SR algorithms.
We argue that it is the low quality of the training datasets that restrict the development of this field. The commonly-used dataset in VFSR is VoxCeleb1~\cite{nagrani2017voxceleb} or VoxCeleb2~\cite{chung2018voxceleb2}. Though the image spatial size in those datasets can reach $800 \times 800$, the contents are blurry and have apparent video compression artifacts, as shown in the top of Fig.~\ref{fig:dataset_comparison}.
Hence, algorithms trained with such datasets will inevitably retain those artifacts and are unable to generate high-quality details.

One may also want to directly apply single-frame face SR methods to videos. However, those approaches always lead to inconsistency among frames, which is a common issue in video applications~\cite{bonneel2015blind,lai2018learning,lei2020blind}.
Many works~\cite{lai2018learning,lei2020blind} have shown that this inconsistency issue could be mitigated by training with multi-frame supervision.
Moreover, exploiting multi-frame information will further improve the restoration performance~\cite{wang2019edvr,chan2021basicvsr}.
Therefore, it is highly desired to have a high-quality VFSR dataset.
Constructing such a high-quality video face dataset is non-trivial work, as there are several complicated steps from the raw videos to the selected high-quality cropped face clips.
In this work, we aim to establish an automatic and  scalable pipeline to collect high-quality face clips from web videos.
Based on this scalable pipeline, we have constructed the Video Face dataset with High Quality (VFHQ), which contains over 16,000 high-fidelity clips of diverse interview scenarios.

\begin{figure}[t]
	\centering
	\includegraphics[width=1\columnwidth]{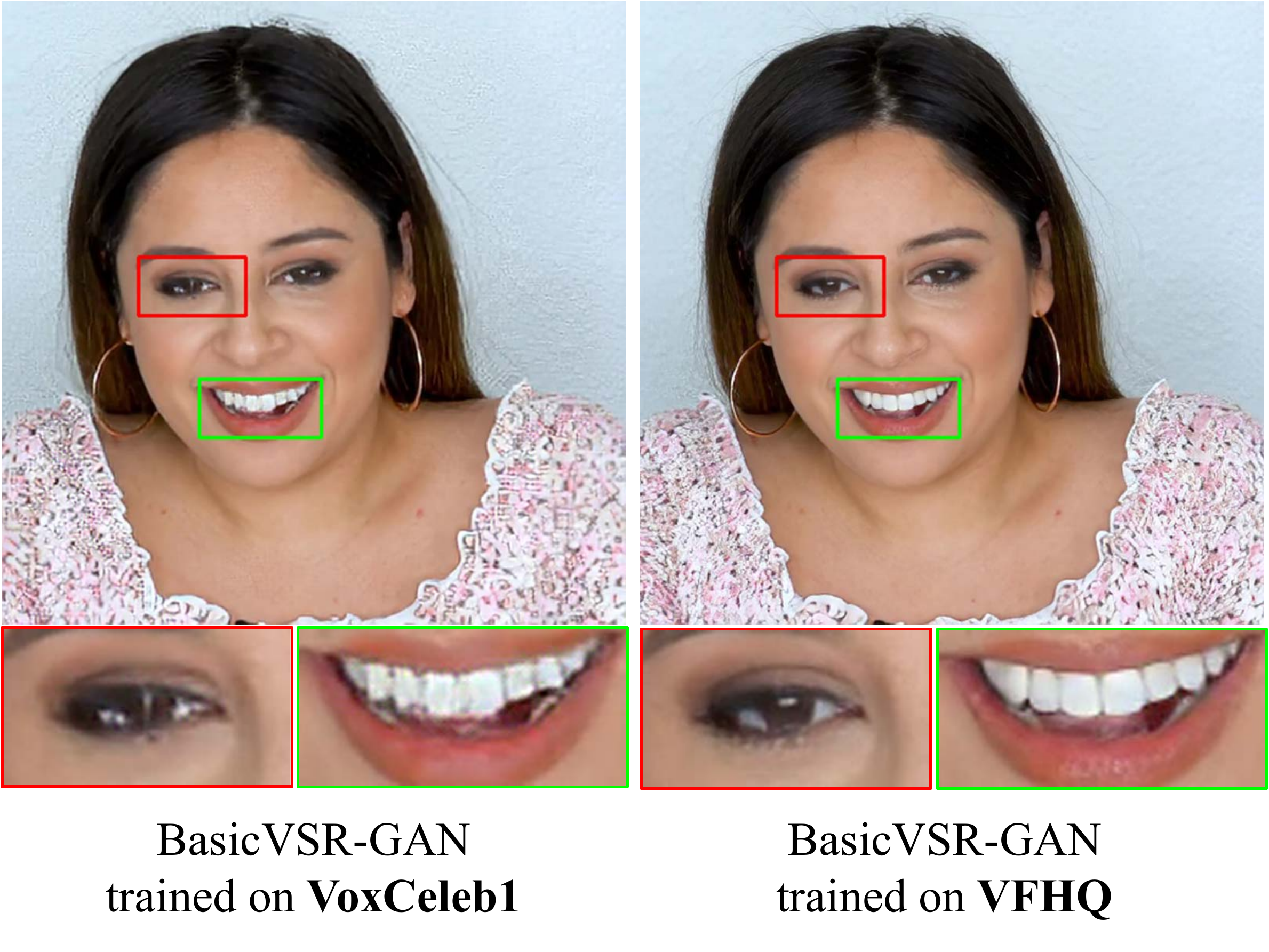}
	\caption{Visual comparison between BasicVSR-GAN models trained with Voxceleb1 and VFHQ dataset, respectively. The high-quality VFHQ dataset helps to recover more visual-pleasing results with finer details.}
	\label{fig:teasor}
	\vspace{-0.4cm}
\end{figure}

It is clear that the quality of VFHQ is superior to VoxCeleb1, as shown at the bottom of Fig.~\ref{fig:dataset_comparison} and Fig.~\ref{fig:statistic} (d). Besides, the clip resolution in VFHQ is between $700 \times 700$ and $1000 \times 1000$, which is close to the resolution of FFHQ images. To verify the necessity of VFHQ compared against VoxCeleb1 for video face SR, we train the BasicVSR~\cite{chan2021basicvsr}, a state-of-the-art video SR method, on these two datasets respectively and compare their results accordingly. Equipped with VFHQ, BasicVSR can achieve more faithful results, and can restore more realistic textures with GAN training~\cite{goodfellow2014generative,ledig2017photo}, as shown in Fig.~\ref{fig:teasor}. We further experimentally show that directly applying single-frame face SR methods trained on FFHQ to restore distorted videos are sub-optimal. Instead, VFHQ not only contains high-fidelity details for each individual frame, but also provides beneficial temporal information to promote better video consistency.

Based on the proposed VFHQ, we further conduct several benchmarking studies on the $\times 4$ bicubic and blind degradation settings. 
We summarize our contributions as follows. \textbf{1)} We establish an automatic and scalable pipeline to collect high-quality face clips from web videos, and construct a high-quality video face dataset VFHQ, which is superior to the commonly-used VoxCeleb1 in both quantitative and qualitative evaluation. \textbf{2)} We further verify the necessity of VFHQ compared against VoxCeleb1 and FFHQ. By using VFHQ, the recent works can achieve better performance than the same models trained on VoxCeleb1. Besides, compared against FFHQ, VFHQ can help to recover more details and mitigate the inconsistency issue in restored videos. \textbf{3)} Based on VFHQ, we evaluate several state-of-the-art methods in both bicubic and blind degradation settings to better understand the potential and limitations of those methods. 

%% file: sections/related_works.tex
\section{Related Work}
\label{sec:related_work}
Face super-resolution is an active problem in computer vision and can be divided into single-frame face super-resolution (SFSR) and  video face super-resolution (VFSR).

This problem has been studied for a long time and please refer to~\cite{jiang2021deep} for a detailed survey.
Different from single image SR methods~\cite{dong2014learning,kim2016accurate,ledig2017photo,timofte2017ntire,haris2018deep,zhang2018image,wang2018esrgan,zhang2020deep,dai2019second,mei2020image,liu2020residual} that directly learn a mapping from low-resolution image to their high-resolution counterpart, most SFSR methods attempt to integrate facial prior knowledge into the CNN architecture. There are three typical types of face-specific priors: geometry priors~\cite{chen2018fsrnet,zhu2016deep,chen2021progressive,yu2018face,xin2019residual}, reference-based priors~\cite{li2018learning,li2020enhanced} and generative facial priors~\cite{wang2021towards,yang2021gan}.
In contrast to the fast development of single-frame face SR, there are few attempts in VFSR~\cite{fang2019self,xin2020video,meishvili2020learning} based on deep neural networks.
All these methods focus on investigating the fusion of spatial and temporal information across frames or the fusion of aural and visual modalities.
They do not consider the facial priors as SFSR does. Therefore, they are similar to general video super-resolution~\cite{xue2019video,chan2021basicvsr,wang2019edvr} except for the used dataset.

The rapid development of the SFSR field can be partly attributed to the richness of image face datasets.
There are several widely-used datasets for training and evaluating the SFSR methods, \textit{e.g.}, Helen~\cite{le2012interactive}, CelebA~\cite{liu2015deep}, LFW~\cite{huang2008labeled}, AFLW~\cite{koestinger2011annotated} and FFHQ~\cite{karras2019style}.
Among them, FFHQ consists of $70,000$ high-quality images whose initial size exceeds $1024 \times 1024$.
Based on the FFHQ dataset, some recent works~\cite{wang2021towards,yang2021gan,chen2021progressive} have achieved superior performance and can restore faces with faithful textures.
Due to the low cost of taking high-definition face pictures and abundant online resources, it is easy to construct such high-quality image datasets without complicated pre-processing.
In contrast to the abundant face image datasets, the most commonly used datasets in VFSR are VoxCeleb1~\cite{nagrani2017voxceleb} and VoxCeleb2~\cite{chung2018voxceleb2}. Although these two datasets contain numerous  utterances of celebrities, the resolution and quality of most videos are so poor that the models trained on these datasets do not have adequate ability to restore high-quality frames as SFSR methods.
In order to fill the gap between the image face dataset and video face dataset, we propose a pipeline to extract high-quality face clips from web videos, and construct a high-quality video face dataset (VFHQ), which could promote the development of the VFSR field.

\begin{figure}[t]
	\centering
	\includegraphics[width=1\columnwidth]{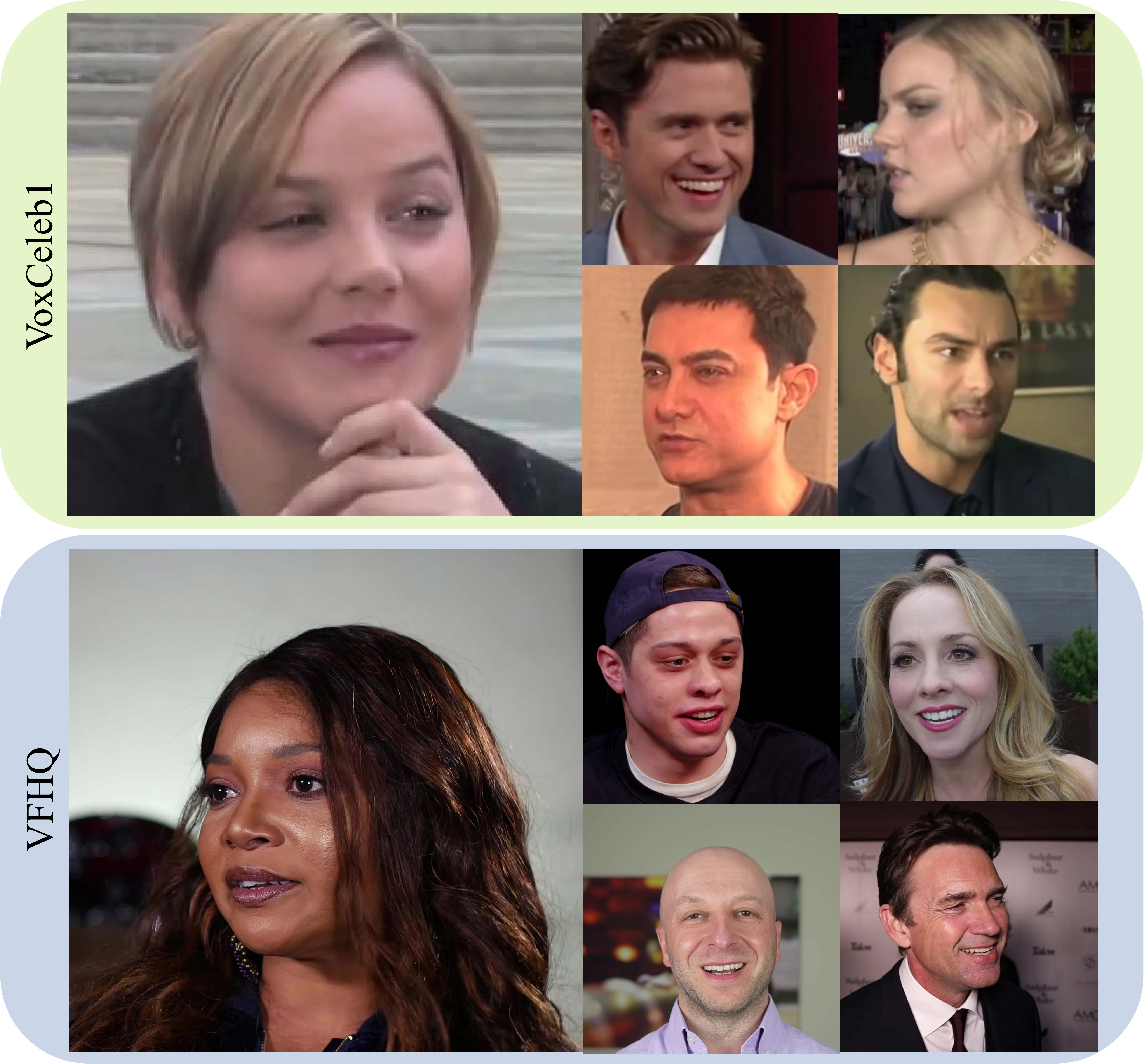}
	\caption{Visual comparisons between the two datasets: VoxCeleb1 (\textbf{top}) and VFHQ (\textbf{bottom}). Images are randomly selected from the dataset. VFHQ images have much higher quality. \textbf{Zoom in for best view}}
	\label{fig:dataset_comparison}
	\vspace{-0.5cm}
\end{figure}

%% file: sections/dataset_description.tex
\section{Dataset Description} \label{sec:data}
\label{sec:dataset_description}

\begin{figure}[t]
	\centering
	\includegraphics[width=1\columnwidth]{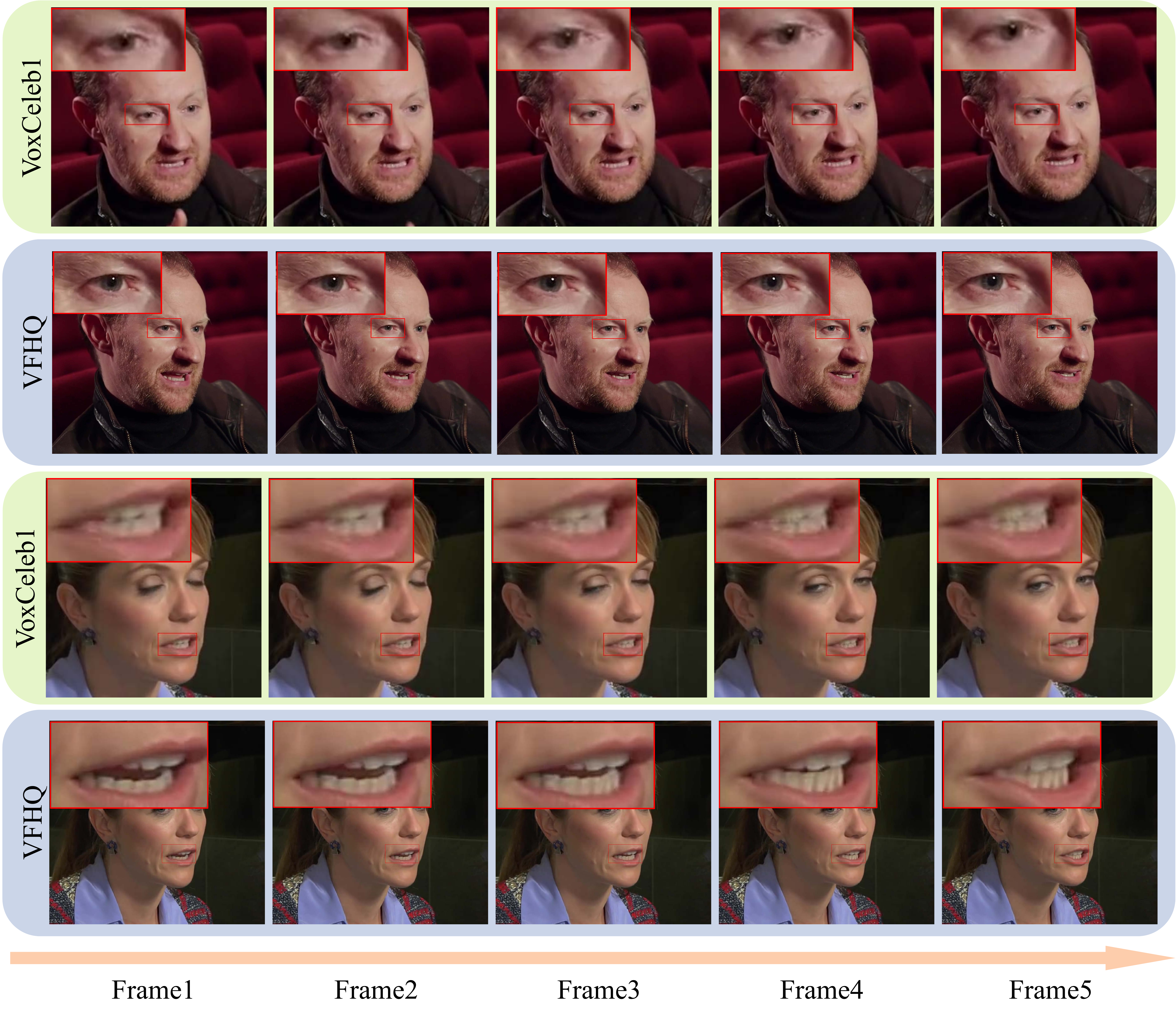}
	\caption{Visual comparison between VoxCeleb1 and VFHQ. For each dataset, we select five consecutive frames and the identity of selected videos is the same. The odd rows are the pictures of VoxCeleb1, while the even rows are the pictures of VFHQ. \textbf{Zoom in for best view}}
	\label{fig:detail_comparison}
	\vspace{-0.3cm}
\end{figure}

Following VoxCeleb1~\cite{nagrani2017voxceleb}, VFHQ is composed of clips for celebrities and is extracted from YouTube videos. The visual comparisons between VoxCeleb1 and VFHQ are shown in Fig.~\ref{fig:dataset_comparison} and Fig.~\ref{fig:detail_comparison}.

The pipeline adopted in collecting VoxCeleb1 and VoxCeleb2 is the same, which means the quality of these two datasets is nearly the same. Here, we only compare VFHQ with VoxCeleb1. 
In Fig.~\ref{fig:dataset_comparison}, we show several images randomly selected from VoxCeleb1. It can be observed that most of the frames in VoxCeleb1 are blurry and of low quality, while the face details in VFHQ are well preserved. We further select two sets of videos with the same identity from these two datasets, and show five consecutive frames within each video, as shown in Fig.~\ref{fig:detail_comparison}. The frames in VFHQ retain relatively high quality across the whole video, while the frames in VoxCeleb1 are distorted with severe compression.

\begin{figure}[t]
	\centering
	\includegraphics[width=1.0\columnwidth]{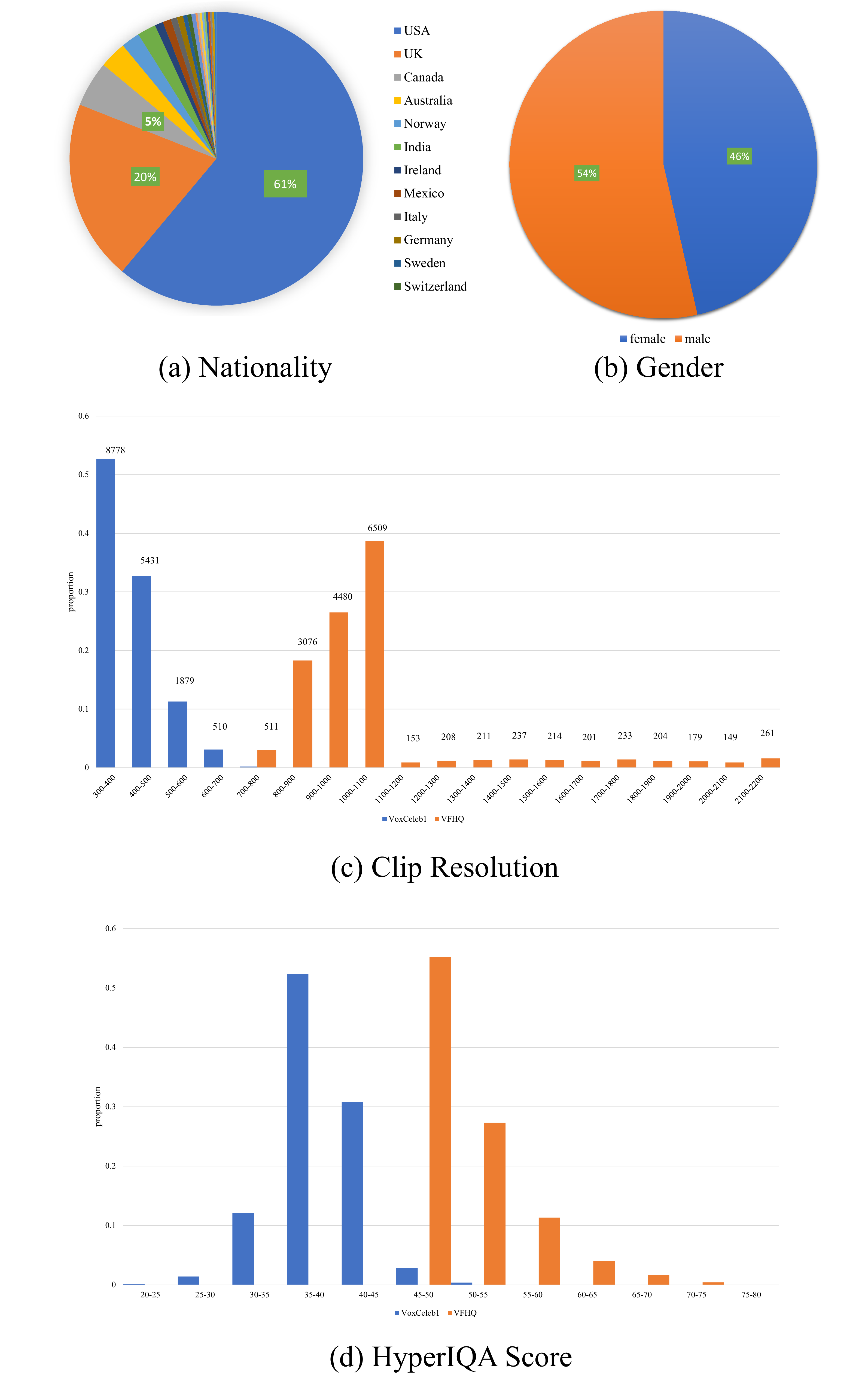}
	\caption{Distribution of the properties of the celebrities in our VFHQ in different aspects. As shown in (a), VFHQ includes persons that come from more than 20 distinct countries. In (b), we notice that the proportion of men and women is roughly the same. The figure (c) demonstrates that the distribution of clip resolution of our VFHQ is different from VoxCeleb1 and the resolution of VFHQ is much higher than VoxCeleb1. Above the bar is the number of clips. Note that we use the length of the shortest side as the clip resolution. The figure (d) shows that the quality of VFHQ is higher than VoxCeleb1 quantitatively.}
	\label{fig:statistic}
	\vspace{-0.3cm}
\end{figure}

Moreover, we present the distribution of VFHQ celebrities in different aspects including nationality and gender. In our VFHQ celebrity list, we include persons that come from more than 20 distinct countries (Fig.~\ref{fig:statistic} (a)). The proportion of men and women is roughly the same (Fig.~\ref{fig:statistic} (b)). Compared with VoxCeleb1, the clip resolution of our VFHQ is much higher (Fig.~\ref{fig:statistic} (c)). The hyperIQA (a blind image quality assessment (BIQA) method for authentically distorted images)~\cite{su2020blindly} score of clips in VoxCeleb1 and VFHQ is shown in Fig.~\ref{fig:statistic} (d), which quantitatively reflects the high-quality of VFHQ.

%% file: sections/dataset_collection.tex
\section{Dataset Collection Pipeline}
\label{sec:dataset_collection}

This section describes our multi-stage approach to collecting the VFHQ dataset, starting from YouTube videos. We adopt several CNN-based algorithms, including face detection (RetinaNet~\cite{deng2019retinaface}), face recognition (ArcFace~\cite{deng2019arcface}), face alignment (AWing~\cite{wang2019adaptive}), tracking (SORT~\cite{bewley2016simple}) and image quality assessment (HyperIQA~\cite{su2020blindly}).

The pipeline involves 1) obtaining the raw videos from YouTube; 2) tracking faces by adopting RetinaNet and SORT algorithms; 3) confirming that the identity of each sub-video is the same by ArcFace; 4) selecting high quality sub-videos (top-three) within each video by calculating the assessment score and face landmark motions. Using this scalable pipeline, we have obtained $16,827$ video clips. We discuss the key stages in the following subsections.

\subsection{Stage 1. Downloading videos from YouTube}
Both VoxCeleb1 and VGGFace2~\cite{cao2018vggface2} provide a name list of celebrity, which contains $1,251$ and $9,131$ celebrities respectively. Based on these two lists, we crawl the corresponding videos from YouTube. Specifically, we append the word `interview 4K' to the name of a celebrity in search query and download the top 20 videos for each celebrity.

\subsection{Stage 2. Face tracking}
For each frame within the video, we first use RetinaNet to detect face bounding boxes and filter out the detections with small sizes (less than $500 \times 500$). Then, all face detections are grouped together into face tracks by SORT. At this stage, we keep the tracks with the frame length between 100 and 2000.

\subsection{Stage 3. Face verification}
Based on the coarse tracks generated by the previous stage, we further refine the tracks to confirm that the detections in each track have the same identity.
This is done by first using ArcFace to extract the feature of each detection and then calculating the $L_{2}$ similarity within every two features. The identities of two frames are considered to be different when the similarity is larger than a threshold $1.24$.
In this case, we will split a long clip into several short clips, in order to make sure that each frame within one short clip belongs to the same identity. At this stage, we also filter out clips that have less than 100 frames.

\subsection{Stage 4. Selecting high-quality clips}
For a clip that has not been filtered out, we are sure that it has a large spatial size (resolution), but we cannot guarantee its quality.
Empirically, we find that HyperIQA~\cite{su2020blindly} owns good generalization ability for face quality assessment in real scenes and we integrate it into our pipeline to help filter out low-quality clips.

Specifically, we first calculate the assessment score (the score of HyperIQA) $AS_{frame}$ of each frame. The score ranges from $0$ to $100$ and the higher value represents better quality. Empirically, we compare it with the threshold $42$. Once the assessment scores of more than four consecutive frames are less than this threshold, we discard these frames and divide the clip into two clips. 
After that, we calculate the average assessment score $AS_{clip}$ of each clip and compare it with the overall threshold (we empirically set it to $45$). The clips of which the average score is less than this threshold are discarded.

After the procedure of the above two steps, we find that some videos have a large number of clips that meet the requirements, while others have only one or two. To increase the dataset diversity and eliminate the imbalance, we finally select top-three high-quality clips for each video by considering both the assessment scores $AS_{clip}$ and landmark motion $M _{clip}$:

\begin{equation}\label{score}
    Score _{clip} = \alpha AS_{clip} + \beta \hat{M}_{clip},
\end{equation}
The landmark motion $M _{clip}$ is calculated on the $98$ landmark points:

\begin{equation}\label{landmark motion}
    M_{clip} = \frac{1}{N \times 98} \sum_{i = 1}^{N-1}  \| \mathcal{L} _{i+1} - \mathcal{L} _{i} \Vert ^{2},
\end{equation}
where $\mathcal{L}_{i}$ and $\mathcal{L}_{i+1}$ are the $98$ landmark results of the $i$ and $i+1$ frames. $N$ is the total frame number in each clip.
We further normalize $M_{clip}$ by:
\begin{equation}\label{motion}
    \hat{M}_{clip} = 0.25M_{clip} + 42.5.
\end{equation}
Empirically, we set $\alpha$ and $\beta$ to $0.5$ and $0.2$, respectively, to balance their importance.

\begin{figure}[t]
	\vspace{-0.6cm}
	\centering
	\includegraphics[width=.85\columnwidth]{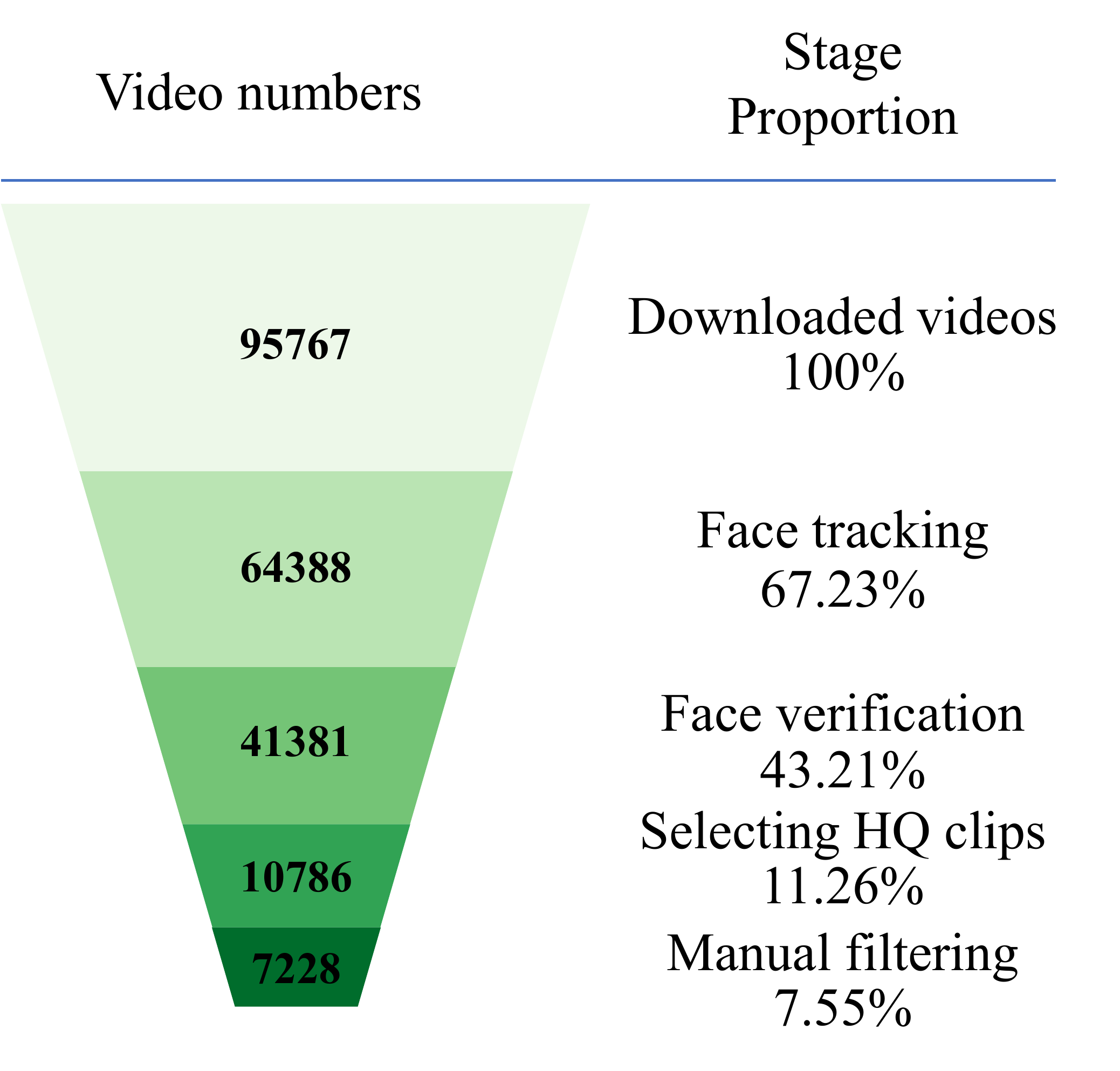} %
	\vspace{-0.3cm}
	\caption{The number and proportion of remained videos after each stage. ``Video number'' means that the number of full videos crawled from YouTube. ``Video clip number'' means that the number of clips split from full videos. VFHQ includes $16,827$ clips from $7,228$ videos.}
	\label{fig:filter_proportion}
	\vspace{-0.3cm}
\end{figure}

\subsection{Stage 5. Manual filtering}

Ideally, the proposed automatic pipeline can filter out all clips with distortion. However, there exists the generalization problem for HyperIQA and we need to manually verify the qualify of remained clips. Compared to directly manual filtering the remained clips (frame by frame) of stage 3, the amount of clips that need to be processed is greatly reduced and the process of verification takes less time. In practice, we uniformly select five frames for each clip and check their quality. The clip will be discarded when all five frames are obviously of low quality.

From stage 2 to stage 5, we discard the unsatisfied videos steadily. Fig.~\ref{fig:filter_proportion} shows the proportion of remaining videos after each stage. We have crawled a total of $95,767$ raw videos and finally obtain $16,827$ clips from $7,228$ videos. 
The percentage of final remaining high-quality videos is about $7.55\%$.

The diversity of motion is considered during collection and VFHQ clips can be categorized into 3 categories according to their motions. We calculate the average pixel displacement of each clip to perform
such division. The ratio of large motion, middle motion and
slow motion are $23.6\%$, $32.2\%$ and $44.2\%$, respectively.

%% file: sections/necessity.tex
\section{The necessity of VFHQ} 
\label{sec:necessity}

The intuitive difference between VFHQ and the other two datasets (i.e, VoxCeleb1 and FFHQ) is that the quality of VFHQ is superior to VoxCeleb1 and FFHQ lacks temporal information. However, the effectiveness of VFHQ is still unclear. 
Hence, we further investigate two questions in the following section.
\begin{enumerate}
	\item \emph{The necessity of our proposed VFHQ compared against VoxCeleb1.} We verify this in the following experiments from two facets: 1) Is VFHQ a more suitable dataset for evaluating existing algorithms? 2) Will training on VFHQ improves the visual quality, for both the MSE-based and GAN-based methods?
	
	\item \emph{The necessity of our proposed VFHQ compared against FFHQ.} We verify this in the following experiments from two aspects: 1) How does the quality of VFHQ compared against FFHQ? 2) Will utilizing the temporal information help to relieve the video consistency issue and further enhance visual quality?
\end{enumerate}

\subsection{Experiment Settings}
We compare different methods on three datasets: FFHQ, VoxCeleb1 and our proposed VFHQ. Specifically, we choose the representative image SR method -- ESRGAN~\cite{wang2018esrgan}, the state-of-the-art video SR method -- BasicVSR~\cite{chan2021basicvsr} and our implemented BasicVSR-GAN. The details of these methods and more experiments with other methods (RRDB, EDVR, EDVR-GAN) can be found in the supplementary materials. 

Recent works~\cite{wang2021towards,yang2021gan,chen2021progressive} are focusing on restoring or generating high-quality faces whose sizes are up to $512 \times 512$. Following GFPGAN~\cite{wang2021towards}, we resize all the images to $512 \times 512$ as HR images. All experiments in this section are performed with a scaling factor of $\times 4$ between LR and HR images/frames. The corresponding LR images/frames are obtained by down-sampling the corresponding HR images/frames with the MATLAB bicubic kernel. 

For better evaluation, we construct two testing datasets, VoxCeleb1-Test and VFHQ-Test. VoxCeleb1-Test contains $20$ sequences that are randomly selected from VoxCeleb2~\cite{chung2018voxceleb2}. VFHQ-Test is composed of $50$ sequences that are randomly selected from VFHQ. Note that these two testing datasets have no overlap with their corresponding training datasets. All other clips that are not included in these two test datasets respectively construct two corresponding training datasets.

\subsection{Comparisons with VoxCeleb1}

\begin{figure}[t]
	\centering
	\includegraphics[width=1\columnwidth]{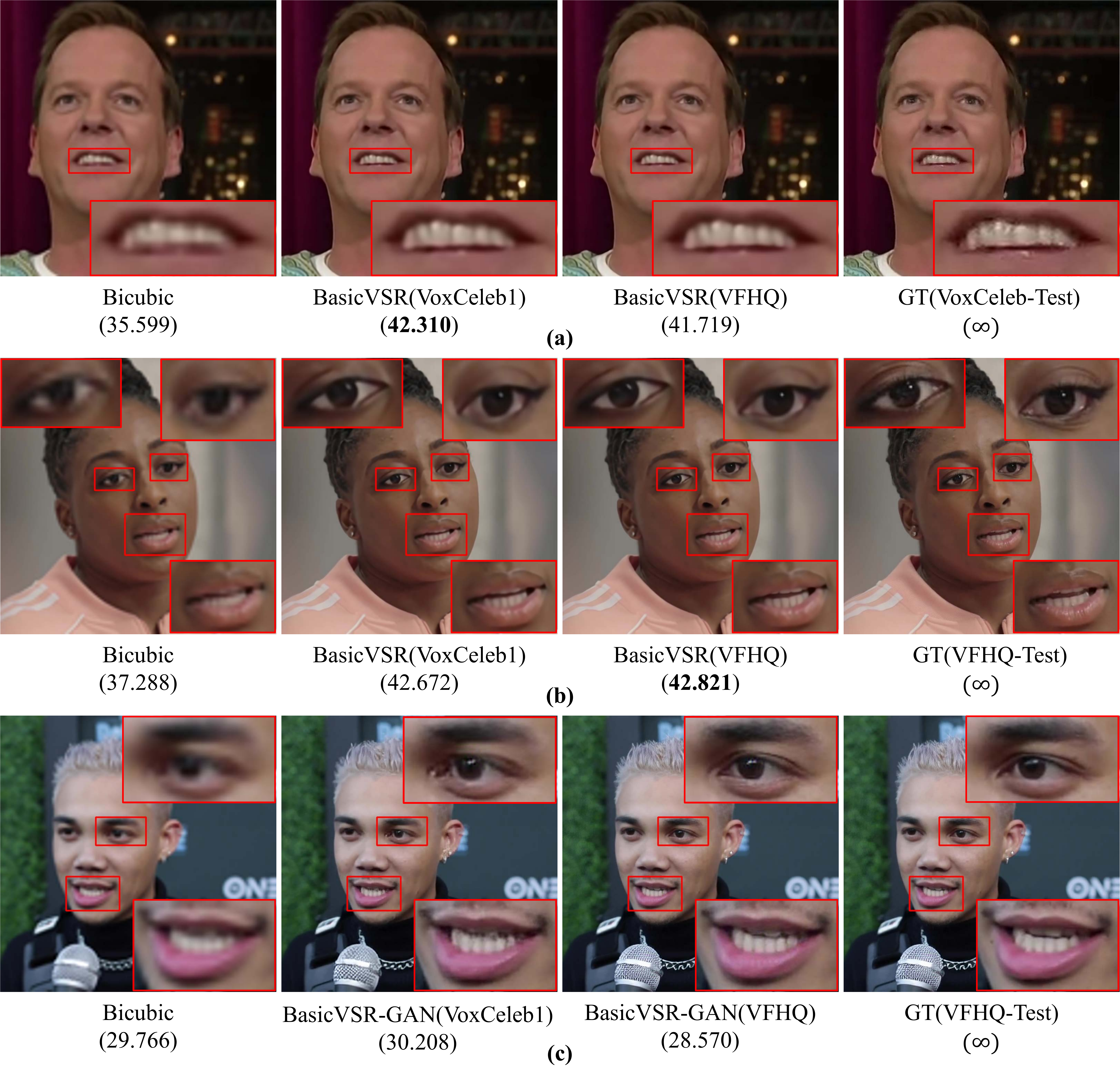} %
	\vspace{-0.5cm}
	\caption{Qualitative comparisons of BasicVSR models trained with VoxCeleb1 and with VFHQ datasets. ((a) evaluated on VoxCeleb-Test, (b) evaluated on VFHQ-Test). (c) Qualitative comparisons of BasicVSR-GAN
	models trained with VoxCeleb1 and with VFHQ datasets. \textbf{Zoom in for best view}}
	\label{fig:necessity_voxceleb}
	\vspace{-0.2cm}
\end{figure}

\begin{figure}[t]
	\centering
	\includegraphics[width=1\columnwidth]{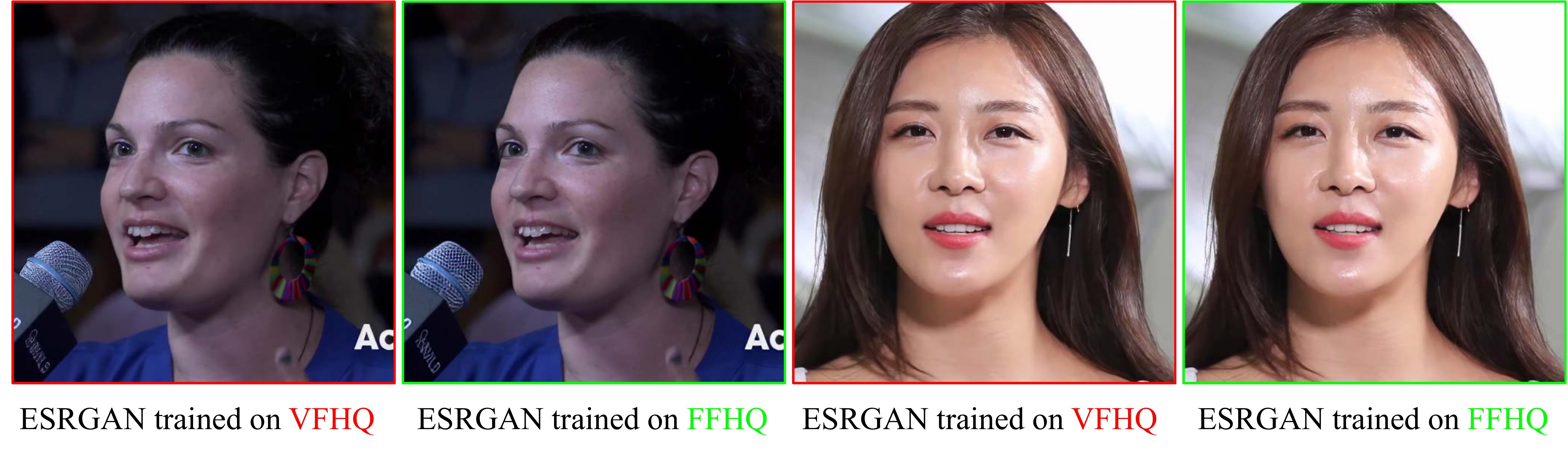} %
	\caption{Visual comparisons between ESRGAN models trained with VFHQ and with FFHQ dataset, respectively. Both models restore similar details in face components. \textbf{Zoom in for best view}}
	\label{fig:vfhq_ffhq}
	\vspace{-0.3cm}
\end{figure}

VFHQ is proposed as a supplement to VoxCeleb1 and we hope VFHQ can be a new dataset for face SR. From Fig.~\ref{fig:dataset_comparison} and Fig.~\ref{fig:statistic}, we can observe that VFHQ is superior to VoxCeleb1 in two facets: image quality and resolution distribution.
To further explore the necessity of VFHQ for face SR, we train BasicVSR based on these two datasets and evaluate their results in VoxCeleb1-Test and VFHQ-Test.

The quantitative evaluation can be found in Tab.~\ref{tab:cmp_datasets}.
Due to the difference in the distribution of VFHQ and VoxCeleb1, for a specific testing dataset, the model trained on the corresponding training dataset achieves the better performance on PSNR/SSIM metrics. However, from visual comparisons in Fig.~\ref{fig:necessity_voxceleb} (a), we can find that: The quality of Ground-Truth (GT) in VoxCeleb1-Test is blurry with distortions. As PSNR is a pixel-wise metric, the restored facial components with lower quality may get a higher value. This phenomenon indicates that VoxCeleb1 dataset is not suitable for making paired test dataset to evaluate the performance of existing methods. Since many works~\cite{fang2019self,xin2020video} evaluate their proposed methods based on the paired test dataset generated by VoxCeleb, we think that a high-quality test dataset to better evaluate existing algorithms is in urgent need. In Fig.~\ref{fig:necessity_voxceleb} (b), it is clear that BasicVSR trained with VFHQ recovers more faithful details in the eyes than BasicVSR trained with VoxCeleb1.

For restoration task, GAN is a common technique for generating more realistic images. Therefore, based on VoxCeleb1 and VFHQ, we further fine-tune their corresponding BasicVSR, obtaining BasicVSR-GAN. As shown in Fig.~\ref{fig:necessity_voxceleb} (c), BasicVSR-GAN trained with VoxCeleb1 fails to retain the fidelity of teeth and tends to generate artifacts, while BasicVSR-GAN trained with VFHQ obtains better teeth shape. Besides, when trained with VFHQ, BasicVSR-GAN is capable of recovering faithful details in the eyes.

In summary, compared against VoxCeleb1, the necessity of VFHQ reflects in two aspects. 1) It is a suitable dataset for evaluating existing algorithms, which can further promote researchers to propose better algorithms with better visual effects. 2) For an algorithm (e.g, BasicVSR), when trained with VFHQ rather than VoxCeleb1, the algorithm can restore more realistic textures. This phenomenon is more obvious when the algorithm is trained with GAN.

\begin{table}[t]
	\centering
	\caption{Quantitative results on VoxCeleb1-Test and VFHQ-Test. Trained on BasicVSR.}
	\label{tab:cmp_datasets}\resizebox{\linewidth}{!}{
		\begin{tabular}{c|c|cc|cc}
			\hline
			\multicolumn{1}{c|}{\multirow{2}{*}{Methods}} & \multicolumn{1}{c|}{\multirow{2}{*}{\begin{tabular}[c]{@{}c@{}}Training\\ Datasets\end{tabular}}} & \multicolumn{2}{c|}{VoxCeleb-Test}                    & \multicolumn{2}{c}{VFHQ-Test}                        \\ \cline{3-6} 
			\multicolumn{1}{c|}{}                                                                             & \multicolumn{1}{c|}{}                         & \multicolumn{1}{c|}{PSNR} & \multicolumn{1}{c|}{SSIM} & \multicolumn{1}{c|}{PSNR} & \multicolumn{1}{c}{SSIM} \\ \hline
			\multirow{2}{*}{BasicVSR}                                                                         & VoxCeleb1                                      & 43.367                    & 0.9829                    & 36.064                    & 0.9410                    \\
			& VFHQ                                  & 42.760                    & 0.9817                    & 36.399                    & 0.9429                    \\ \hline
		\end{tabular}
	}
	\vspace{-0.2cm}
\end{table}

\begin{table}[t]
	\centering
	\caption{Quantitative results with different training input frames for BasicVSR. Evaluated on VFHQ-Test.``Length'' indicates the input frame length of network during the training phase.}
	\label{tab:basicvsr_multi_vs_single}\scalebox{0.90}{
		\begin{tabular}{c|c|c|c}
			\hline
			Method                    & Length & PSNR (dB) & SSIM   \\ \hline
			\multirow{2}{*}{BasicVSR} & L=1    & 35.213   & 0.9293 \\ \cline{2-4} 
			& L=7    & 36.258 (+1.045)   & 0.9412 (+0.0119) \\ \hline
	\end{tabular}}
	\vspace{-0.5cm}
\end{table}

\begin{figure}[t]
	\centering
	\includegraphics[width=1\columnwidth]{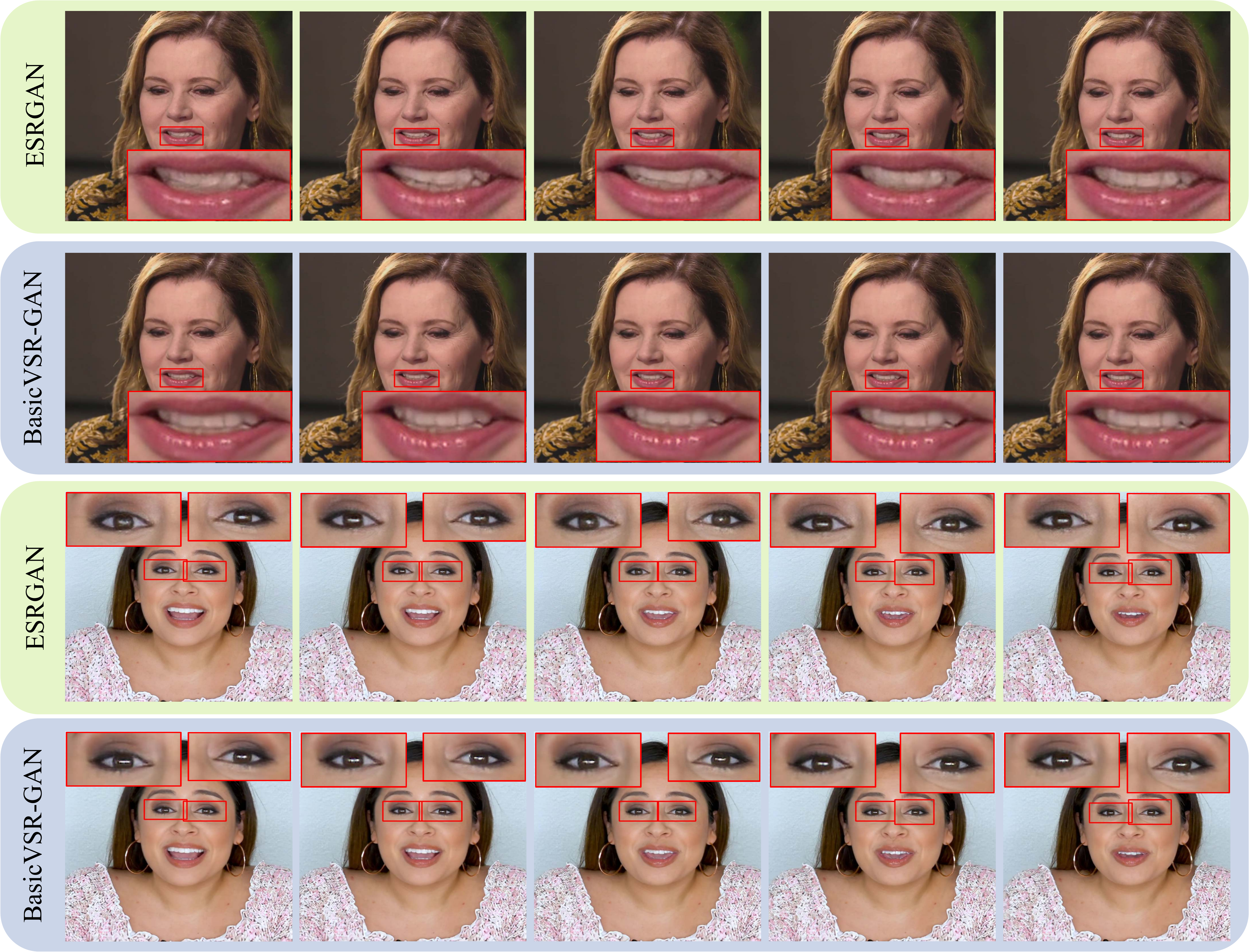} %
	\caption{Qualitative comparisons between BasicVSR-GAN and ESRGAN. We present the result of five consecutive frames. \textbf{Top}: BasicVSR-GAN can restore the complete tooth shapes while ESRGAN mixes all the teeth together.  \textbf{Bottom}: the bright spots in eyes keep changing for different frames in ESRGAN, while are consistent in BasicVSR-GAN. \textbf{Zoom in for best view}}
	\label{fig:esrgan_vs_basicvsr-gan}
	\vspace{-0.2cm}
\end{figure}

\begin{table*}[t]
	\centering
	\caption{Benchmarking results with \textbf{bicubic} degradation model (evaluated on VFHQ-Test). Average PSNR/SSIM values for scaling factor $\times 4$. \redbf{Red} and \blueud{blue} indicates the best and second best performance. The sampling interval in the testing phase is equal to $5$.}
	\label{tab:bicubic-comparison}{
		\begin{tabular}{c|c|cccc|ccc}
			\hline
			\multicolumn{1}{c|}{\multirow{2}{*}{Interval}} & \multicolumn{1}{c|}{\multirow{2}{*}{Metrics}} & \multicolumn{4}{c|}{MSE-based}                                                                                        & \multicolumn{3}{c}{GAN-based}                                                                   \\ \cline{3-9} 
			\multicolumn{1}{c|}{}                          & \multicolumn{1}{c|}{}                         & \multicolumn{1}{c}{Bicubic} & \multicolumn{1}{c}{RRDB} & \multicolumn{1}{c}{EDVRM} & \multicolumn{1}{c|}{BasicVSR} & \multicolumn{1}{c}{ESRGAN} & \multicolumn{1}{c}{EDVRM-GAN} & \multicolumn{1}{c}{BasicVSR-GAN} \\ \hline
			\multirow{2}{*}{5}                              & PSNR                             &    31.964                          &         35.332                 &    \blueud{36.090}                        &          \redbf{36.258}               &    32.803                         &         33.592                 &    32.327                               \\
			& SSIM                                                                      &       0.8939                   &    0.9302                       
			&       \blueud{0.9399}                   &    \redbf{0.9412}                           &       0.8961                   &    0.9089                            &       0.8869             \\ \hline              
	\end{tabular}}
\vspace{-0.2cm}
\end{table*}

\begin{figure*}[t]
	\centering
	\includegraphics[width=\linewidth]{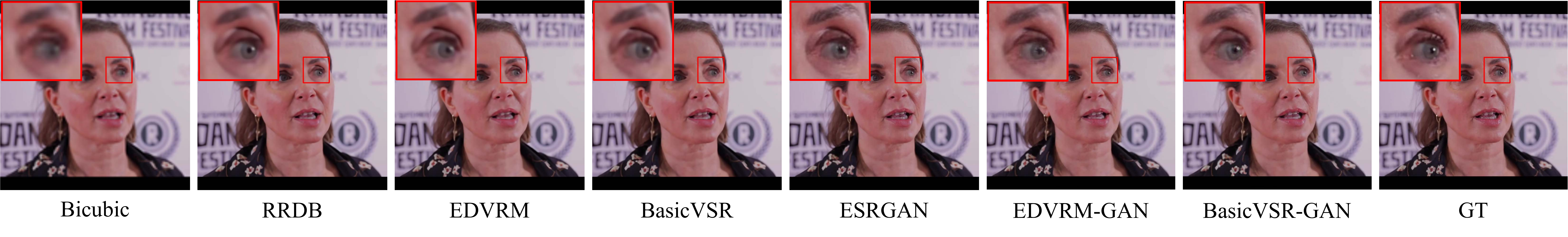}
	\vspace{-0.7cm}
	\caption{Qualitative comparisons by different models in $\times 4$ bicubic degradation setting. \textbf{Zoom in for best view}}
	\label{fig:bicubic_benchmark}
	\vspace{-0.2cm}
\end{figure*}

\subsection{Comparisons with FFHQ}
FFHQ~\cite{karras2019style} is a high-quality image dataset of human faces. Since both VFHQ and FFHQ are high-definition face datasets, we wonder how does the quality of VFHQ compare to FFHQ. To clarify this, we trained two ESRGAN models with VFHQ and with FFHQ datasets, respectively. The visual comparisons show in Fig.~\ref{fig:vfhq_ffhq}. We can observe that the ESRGAN model trained with VFHQ dataset can restore similar details in face components as the ESRGAN model trained on FFHQ dataset, reflecting the high-quality of our collected VFHQ. 

In many scenarios that require face SR, directly applying single-frame face SR methods is an option.
However, it is a sub-optimal option since it ignores the temporal information in the videos.
By utilizing temporal information, there are two benefits. 1) It helps achieve better results by considering complementary information between adjacent frames. 2) It can mitigate inconsistency issues in restored videos.

To clarify this, we apply ESRGAN~\cite{wang2018esrgan} trained with FFHQ~\cite{karras2019style} to VFHQ-Test and compare the results with BasicVSR-GAN, as shown in Fig.~\ref{fig:esrgan_vs_basicvsr-gan}.
We can draw the following observations.
\textbf{1)} For facial components like teeth ($1$st and $2$nd row), BasicVSR-GAN can restore the complete tooth shapes while ESRGAN mixes all the teeth together.
\textbf{2)} Since ESRGAN is a single-frame method and do not consider the information among consecutive frames, each frame in the restored video is independent of each other. Although the motion contained in these frames are small, ESRGAN still leads to obviously pixel jittering. Specifically, the shape of teeth (top) and the bright spots in eyes (bottom) keep changing among the restored five frames. \textbf{3)} Exploiting temporal information in VFHQ is effective to eliminate the inconsistency and improve both the qualitative and quantitative performances. 

We also conduct comparisons under the same network architectures to validate the effectiveness of temporal information in improving performance. As shown in Tab~\ref{tab:basicvsr_multi_vs_single}, on the VFHQ-Test dataset, BasicVSR with seven input frame length (L=7) outperforms
BasicVSR with only one frame (L=1, with equivalent computation FLOPs) by a large margin. It indicates that multi-frame temporal information is pivotal for improving the restoration performance of face videos.

In summary, the quality of VFHQ is comparable with FFHQ. For restoring distorted videos (especially with large motion), compared to FFHQ, the temporal information in VFHQ is pivotal for relieving the video consistency issue and improving the visual quality of restored videos.

%% file: sections/benchmark.tex
\begin{table*}[]
	\centering
	\caption{Benchmarking results with \textbf{blind} degradation model (evaluated on VFHQ-Test). Average PSNR/SSIM/LPIPS values for scaling factor $\times 4$. \redbf{Red} and \blueud{blue} indicates the best and second best performance. The sampling interval in the testing phase is equal to $5$.}
	\setlength{\tabcolsep}{1mm}
	\label{tab:blind-comparison}{
		\begin{tabular}{c|c|ccc|ccc|cc}
			\hline
			\multicolumn{1}{c|}{\multirow{2}{*}{Interval}} & \multicolumn{1}{c|}{\multirow{2}{*}{Metrics}} & \multicolumn{3}{c|}{MSE-based}                                                            & \multicolumn{3}{c|}{GAN-based}                                                                   & \multicolumn{2}{c}{GAN-prior based}                    \\ \cline{3-10} 
			\multicolumn{1}{c|}{}                          & \multicolumn{1}{c|}{}                         & \multicolumn{1}{c}{Bicubic} & \multicolumn{1}{c}{EDVRM} & \multicolumn{1}{c|}{BasicVSR} & \multicolumn{1}{c}{EDVRM-GAN} & \multicolumn{1}{c}{BasicVSR-GAN} & \multicolumn{1}{c|}{DFDNet} & \multicolumn{1}{c}{GFPGAN} & \multicolumn{1}{c}{GPEN} \\ \hline
			\multirow{3}{*}{5}            & PSNR                         
			&     26.842      &   \blueud{29.457}                
			&    \redbf{29.472}      &   26.682        
			&     25.813      &   25.178                         
			&     25.978      &   26.672                    \\
			& SSIM                                          
			&     0.7909      &   \blueud{0.8428}               
			&     \redbf{0.8430}      &    0.7638                           
			&     0.741       &    0.7560                         
			&     0.7723      &      0.7768 \\
			& LPIPS  & 0.4098 & 0.3288 & 0.3309 & \redbf{0.3076} & \blueud{0.3214} & 0.4008 & 0.3446 & 0.3607 \\
			\hline             
	\end{tabular}}
\vspace{-0.2cm}
\end{table*}

\section{Benchmark Experiments}
\label{sec:benchmark}
\subsection{Degradations} To comprehensively evaluate existing methods on VFHQ, we select two degradation models, the bicubic degradation model and the blind degradation model. The first one is classical in super-resolution and the second one is closer to real-world degradation. Details of these two degradations are described as follows.

\noindent \textbf{Bicubic degradation} model is implemented by adopting the Matlab function \textit{imresize}. The downsample scale is $\times 4$.

\noindent \textbf{Blind degradation} model~\cite{wang2021towards,li2020blind} is implemented by following the practice in ~\cite{wang2021towards}.
Considering the compression type in image and video datasets is different, we use \textit{FFMPEG} rather than JPEG to simulate the compression. To be specific, the degradation model is:

\begin{equation} \label{blind}
	\boldsymbol{x}=\left[\left(\boldsymbol{y} \circledast \boldsymbol{k}_{\sigma}\right) \downarrow_{r}+\boldsymbol{n}_{\delta}\right]_{\mathrm{FFMPEG}_{\mathrm{crf}}}
\end{equation}
where $\boldsymbol{x}$ and $\boldsymbol{y}$ are paired low-resolution and high-resolution clips. The $\boldsymbol{k}_{\sigma}$, $r$, $\boldsymbol{n}_{\delta}$ and $\mathrm{crf}$ are Gaussian blur kernel, down-sampler factor, additive white Gaussian noise, constant rate factor (decides how many bits will be used for each frame), respectively. The $r$ equals $4$ in experiments. The sampling range of ${\sigma}$, ${\delta}$ and $\mathrm{crf}$ are $\{0.1: 10\}$, $\{0: 10\}$, $\{18: 25\}$, respectively. Note that for each individual training pair, we only sample one value for ${\sigma}$ and $\mathrm{crf}$, whereas the ${\delta}$ varies among frames in the clip by following~\cite{tassano2020fastdvdnet}.

\subsection{Comparison in Bicubic Degradation}

We conduct experiments with the MSE-based and GAN-based methods. Specifically, for MSE-based methods, we select RRDB~\cite{wang2018esrgan}, EDVRM~\cite{wang2019edvr}, BasicVSR~\cite{chan2021basicvsr}. For GAN-based methods, we select ESRGAN~\cite{wang2018esrgan} and EDVRM-GAN and BasicVSR-GAN, which are fine-tuned based on their corresponding PSNR-oriented models with generative adversarial loss. In the training phase, to increase the motion range, we interval sample a continuous video and input the newly composed video to the network. To be specifically, the sampling interval is $\{3: 7\}$.  
In the testing phase, we also evaluate the performance of difference sampling intervals of test datasets. Here we only show the results whose sampling interval is equal to $5$. Results of other intervals and more visual comparison among these methods can be found in the supplementary materials. 

Tab.~\ref{tab:bicubic-comparison} shows a quantitative comparison between these methods. Consistent with the performance in the general video super-resolution field, BasisVSR achieves the best performance in PSNR and SSIM metrics. The visual comparison of these methods is shown in Fig.~\ref{fig:bicubic_benchmark}, for the current test image, we can find that ESRGAN, EDVRM-GAN and BasicVSR-GAN can restore faithful facial details. This indicates that for video face super-resolution task, specifically in $\times 4$ bicubic degradation setting, current methods are capable of restoring high-quality face videos. In a larger scale ratio (e.g, $\times 8$), the performance gap between these methods is larger and there needs more investigation for a larger scale ratio in the bicubic setting. Experiments for $\times 8$ scale can be found in the supplementary materials.

\subsection{Comparison in Blind Degradation}

\begin{table}[t]
	\centering
	\caption{Quantitative results of combining MSE-based method and GAN-prior based method. Evaluated on VFHQ-Test. The sampling interval in the testing phase is equal to $5$.}
	\label{tab:mse_gan_prior}\scalebox{0.80}{
	\begin{tabular}{c|c|c|c}
		\hline
		Interval           & Metrics & EDVRM+GFPGAN & BasicVSR+GFPGAN \\ \hline
		\multirow{3}{*}{5} & PSNR    & 27.879       & 27.868          \\ \cline{2-4} 
		& SSIM    & 0.8198       & 0.8195          \\ \cline{2-4} 
		& LPIPS   & 0.3265       & 0.3266          \\ \hline
	\end{tabular}}
\vspace{-0.5cm}
\end{table}

Similar to the benchmarking study conducted in the bicubic degradation setting, we evaluate the MSE-based and GAN-based methods in the blind degradation setting. Considering that recent GAN-prior based methods~\cite{wang2021towards,yang2021gan} and DFDNet~\cite{li2020blind} can restore realistic faces on both synthetic and real-world datasets, we also include those methods for comparison. Here, the testing datasets are synthesized based on the same degradation model used in the training pairs. For these three algorithms, we directly apply their released pre-trained models to distorted videos. We also show the restored results of which the sampling interval is equal to $5$.

The quantitative results are listed in Tab.~\ref{tab:blind-comparison}. We find that in the blind degradation setting, the gap between EDVR and BasicVSR on PSNR/SSIM metrics is smaller than the bicubic degradation. For LPIPS metric, we only evaluate the performance of five frames within each restored sequence and EDVRM-GAN achieves the best performance among these methods. 

The strategy of applying both MSE-based method and GAN-prior method to restore distorted sequence is also adopted and the results are listed in Tab~\ref{tab:mse_gan_prior}. Although the performance of EDVRM+GFPGAN and BasicVSR+GFPGAN is better than GFPGAN on LPIPS metric, their performance is inferior to their corresponding GAN-based methods. It indicates that end-to-end training is a better strategy. The design of combining MSE-based methods and GAN-prior based methods into a unified network is left as our future work.

\begin{figure}[t]
	\centering
	\includegraphics[width=1\columnwidth]{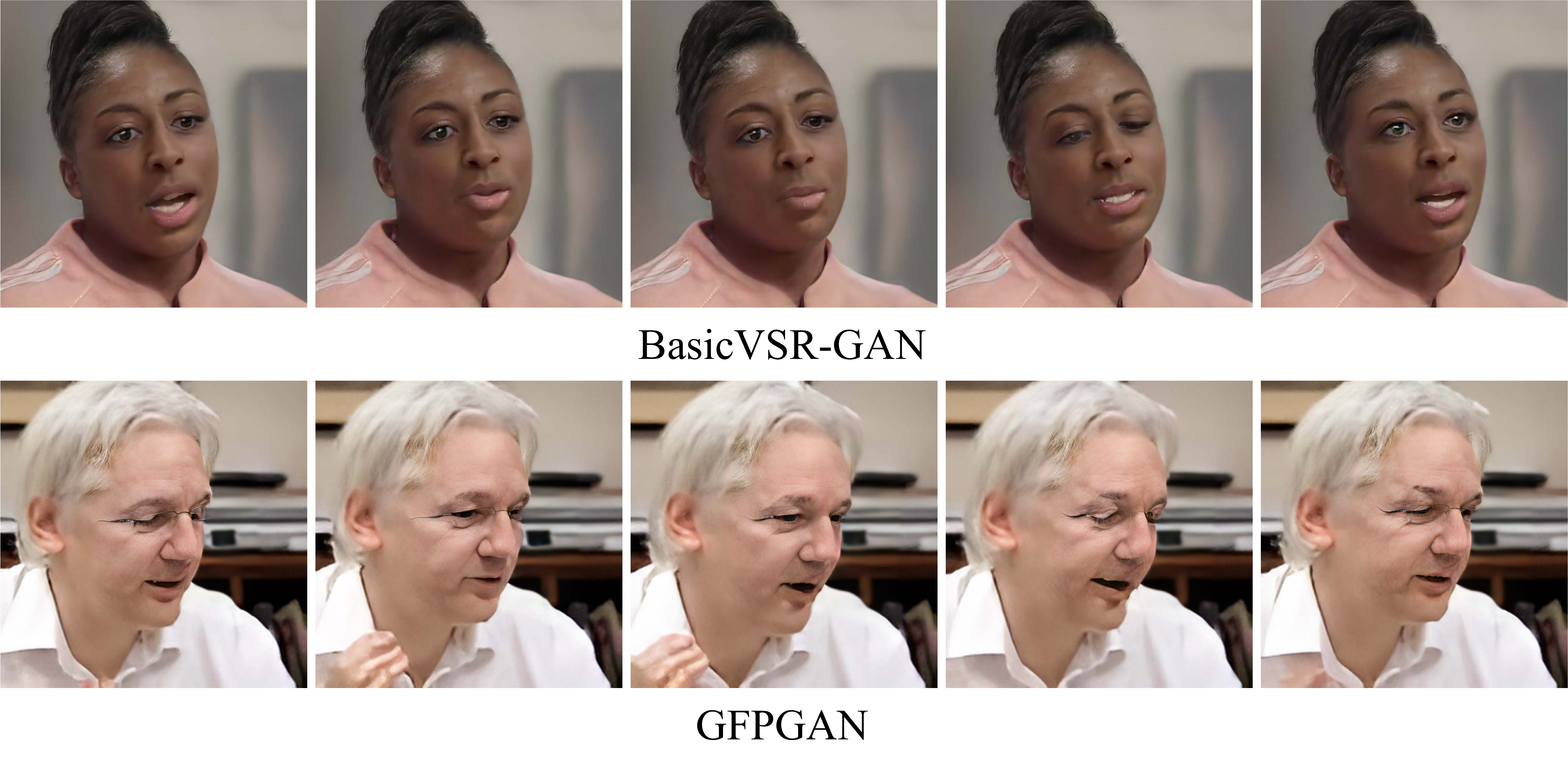} %
	\caption{Limitations of BasicVSR-GAN and GFPGAN.}
	\label{fig:blind_benchmark}
	\vspace{-0.4cm}
\end{figure}

Unlike the bicubic degradation setting, existing methods have limitations in the blind setting, as shown in Fig.~\ref{fig:blind_benchmark}. Specifically, for BasicVSR-GAN, although with end-to-end training, it can not restore realistic faces when the degradation of the input video is relatively severe (still in the range of training data distribution). For GFPGAN, it produces unnatural results for very large poses. Since it only takes the corresponding distorted face as the input, there exists obvious inconsistency in the restored videos. More visual results are shown in the supplementary materials.

